\documentclass[conference]{IEEEtran}

\makeatletter
\def\ps@headings{%
\def\@oddhead{\mbox{}\scriptsize\rightmark \hfil \thepage}%
\def\@evenhead{\scriptsize\thepage \hfil \leftmark\mbox{}}%
\def\@oddfoot{}%
\def\@evenfoot{}}
\makeatother
\usepackage{graphicx}

\usepackage{flushend}

\begin{document}
\title{A Protocol for a Secure Remote Keyless Entry System Applicable in Vehicles using Symmetric-Key Cryptography}

\author{\IEEEauthorblockN{Tobias Glocker}
\IEEEauthorblockA{University of Vaasa\\
Email: tglo@uva.fi}
\and
\IEEEauthorblockN{Timo Mantere}
\IEEEauthorblockA{University of Vaasa\\
Email: timan@uva.fi}
\and
\IEEEauthorblockN{Mohammed Elmusrati}
\IEEEauthorblockA{University of Vaasa\\
Email: moel@uva.fi}}

\maketitle

\begin{abstract}
In our modern society comfort became a standard. This comfort, especially in cars can only
be achieved by equipping the car with more electronic devices. Some of the electronic devices
must cooperate with each other and thus they require a communication channel, which can be wired
or wireless. In these days, it would be hard to sell a new car operating with traditional keys.
Almost all modern cars can be locked or unlocked with a Remote Keyless System. A Remote Keyless
System consists of a key fob that communicates wirelessly with the car transceiver that is responsible
for locking and unlocking the car. However there are several threats for wireless communication
channels.

This paper describes the possible attacks against a Remote Keyless System and introduces a secure protocol
as well as a lightweight Symmetric Encryption Algorithm for a Remote Keyless Entry System applicable in vehicles.\\
\end{abstract}

 \IEEEpeerreviewmaketitle

\section{Introduction}
A Remote Keyless Entry System consists of a key fob and a car transceiver that is responsible for locking and
unlocking the car. Instead of locking or unlocking the car with a traditional key the user presses
a button on the key fob to lock or unlock the car. Unfortunately the keyless cars are  "increasingly
targeted by thieves" [1]. Criminals steal vehicles "through the re-programming of remote-entry keys".
Thus, some insurance companies have denied the insurance for this issue. In addition a Remote Keyless
System is vulnerable against a Scan Attack, Playback Attack, Two-Thief Attack, Challenge Forward
Prediction Attack and a Dictionary Attack. Another threat for Remote Keyless Entry Systems are On-Board-Diagnose (OBD)
key programmers.

A Scan Attack is critical for systems which use the rolling code technique [2]. It can be performed against such
systems by sending different codes to the car transceiver as long as the sent code matches with the code of the
car transceiver. This type of the attack is the simplest one. How long it takes to unlock the car with this attack "depends
on the number of bits in the random challenge, the random challenge-generation method, and the number of trials conducted
by the intruder" [3].

Another possible attack against Remote Keyless Entry Systems is the so called Playback Attack [3]. Here, an
intruder has a device which is capable of recording messages sent wirelessly. Later on, when the car driver is away, the intruder
can send the recorded messages to the car transceiver to unlock the car.

The Two-Thief Attack [3] is the most known attack. In a passive Keyless Entry System one thief stands next to the car while the other
one stands next to the car owner. Assume that the car owner is hundred meters away from the car. Both thieves use devices to amplify
signals. The thief standing next to the car pulls the door handle. By pulling the door handle, the car transceiver sends an interrogation message to the 
Customer-Identification Device (CID) [4] which is like a key fob or a credit card, kept in the car owner's pocket. Since the CID is outside
the transmission range, the amplifier of the thief standing next to the car amplifies the signal so that it can be received by the amplifier
of the thief that stands next to the car owner. From there it will be forwarded to the CID. The CID responds with a valid code which will be
transmitted to the car transceiver over the amplifier devices of the thieves.

In the Challenge Forward Prediction Attack the intruder has a device that records several interrogation messages that are sent from the car
transceiver when the door handle is pulled. Based on the recorded interrogation messages the intruder tries to predict the next one. Then the
intruder can go to the car owner and send the predicted interrogation message to the CID located in the car owner's pocket. The CID responds
with a message which will be recorded. Afterwards, the intruder goes back to the car, pulls the door handle and plays back the recorded message
from the CID.

Cars can also get unlocked with copied car keys. There are OBD key programmers that can be used to copy car keys [5]. Assume that a thief
goes to a car shop and asks for a test drive. During the test drive the thief stops, plugs the key programming device in the OBD and copies all
key information to a non-programmed key fob. After the test drive the thief returns the car with the original car key. In the night the thief goes
again to the car shop and steals the car with the copied key fob.

A further threat are jammers [6]. Jammers are devices that emit signals in the same frequency range as key fobs to create a strong interference
that blocks the communication between key fob and car transceiver. When the driver leaves the car and presses the lock button on the key fob, the 
car will not lock if there is an active jammer in the range of 30m. Although, in most of the cars the indicators blink two times after the car has been locked, some people do not pay any attention to it. After the driver has left the parking yard, the thief can open the unlocked car and steal it by accessing the OBD 
interface.

\section{Functionality of the proposed Protocol for a Secure Remote Keyless Entry System}
To implement this Lightweight Symmetric Algorithm for a Remote Keyless Entry System the following 
requirements must be fulfilled. First it is important that the microcontroller used in the key fob and in the
board computer has an Electrical Erasable Programmable Read-Only Memory (EEPROM) with 4kB to store 2000 numbers
with the size of 2 Bytes. Furthermore a strong Random Number Generator should be used to avoid the prediction
of random numbers. Another important issue is the power consumption of the microcontroller and the Radio Frequency (RF)
unit inside the key fob because they are battery-operated. 

\begin{figure}[!h]
\centering
\includegraphics[width=2.5in]{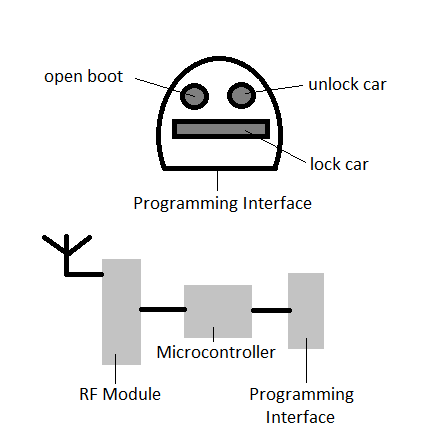}
\caption{Proposed Key Fob with Main Components.}
\label{fig_keyfob}
\end{figure}

The car must also have a board computer, equipped with a car transceiver, for generating random numbers and for sending them to the key
fob. In addition the board computer and the car transceiver must be connected to an accurate current measurement, capable of detecting fluctuations
from one millisecond to another, in order to generate random numbers. Furthermore, the key fob should also be able
to generate random numbers. It is important that Random Number Generators (RNGs) are used but not Pseudo Random Number Generators (PRNGs) [7]. PRNGs generate random numbers whose sequence will be repeated after some time while for RNGs, the probability of repeated random number sequences is vanishingly small. It is to mention that the proposed key fob contains a button for locking the door, unlocking the door, opening the boot and a programming interface (see Fig. 1). In the central console there must be two interfaces through that the key fobs can communicate with the board computer. At certain time intervals the driver of the car will be informed to update the key fobs. During the updating process the board computer will generate new 
random numbers and write them to the memory of the key fobs and the car transceiver. The updating process works only, if both key fobs
are connected to the board computer.

When the car is delivered, the buyer gets two key fobs. The board computer as well as the key fobs are pre-programmed.

\begin{figure}[!h]
\centering
\includegraphics[width=3.3in]{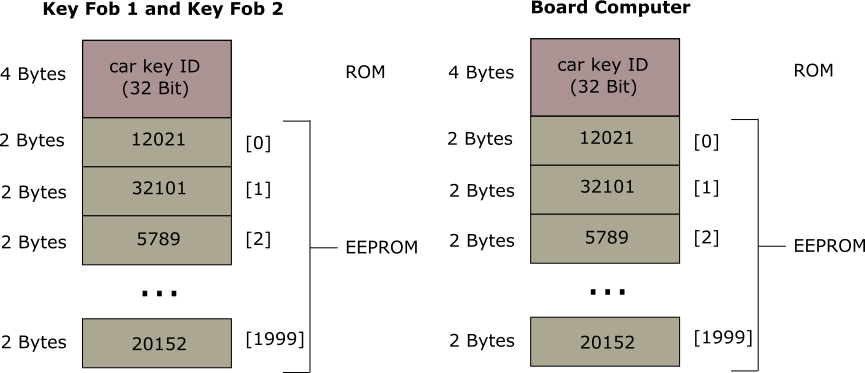}
\caption{Memory Layout of the Key Fob and the Car Transceiver.}
\label{fig_MemLayout}
\end{figure}

Fig. 2 represents the memory layout of the key fob and the car transceiver. The size of the EEPROM is 4kB to store 2000 numbers in 
the range between 0 and 65535. The car key ID is stored in a Read-Only Memory (ROM). It is a unique ID that is assigned to each car. 

In Fig. 3 a successful transaction for sending an instruction command from the key fob to the car transceiver is illustrated.
When the driver presses a button on the key fob, the key fob sends an encrypted car key ID to the car transceiver.
There, the received car key ID will be compared with the one stored in the memory of the car transceiver. If the comparison was
successful, the car transceiver generates ten random numbers in the range between 0 and 1999. These ten
random numbers are sent to the key fob. The first five random numbers are used as indices for the memory locations containing the
numbers used to build a key, while the last five random numbers are used as indices of the memory locations whose values are needed for the encryption (see Fig. 5). The first five numbers of the requested memory locations are sent encrypted and appended to the car transceiver. When the car transceiver receives the so called authentication message then it reads the requested memory locations from its own memory, encrypts and appends them to build its own authentication message. Then it compares the own authentication message with the received one. 

\begin{figure}[!ht]
\centering
\includegraphics[width=3.1in]{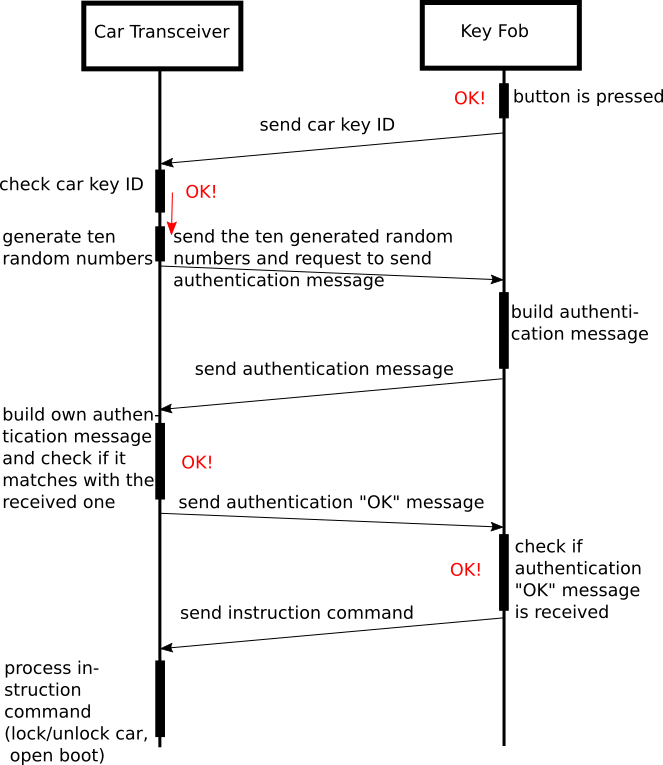}
\caption{Successful Transaction for sending an instruction command from Key \noindent\hspace*{8mm} Fob  to car transceiver.}
\label{fig_Transaction_1}
\end{figure}

\begin{figure}[!ht]
\centering
\includegraphics[width=3.3in]{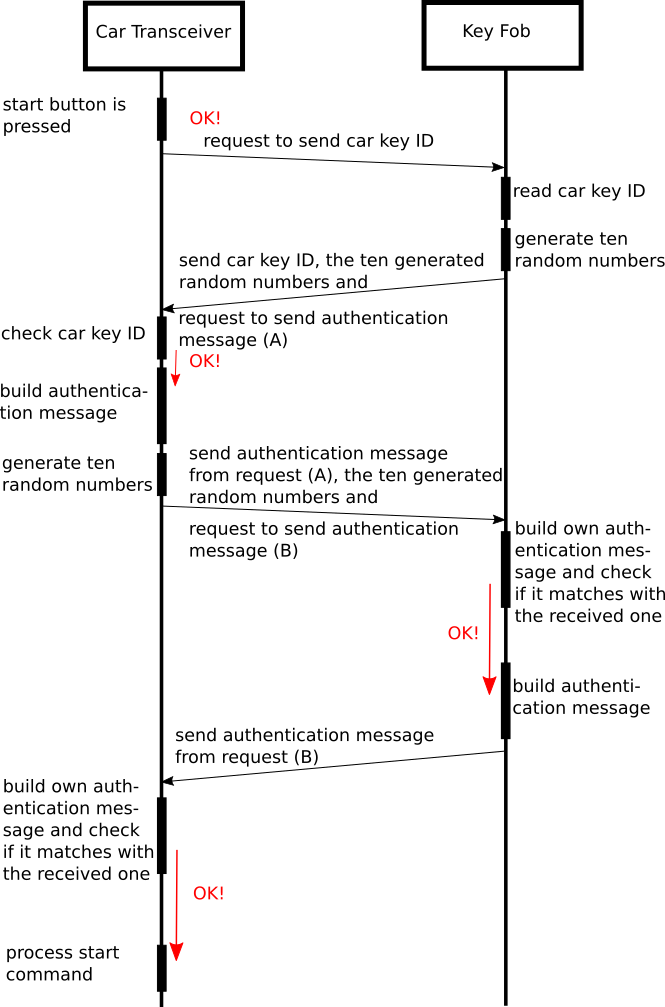}
\caption{Successful Transaction between car transceiver and Key Fob  for \noindent\hspace*{10mm} starting the car.}
\label{fig_Transaction_2}
\end{figure}

If the comparison was successful, the car transceiver sends an authentication "OK" message to the key fob. After receiving the authentication "OK" message, the key fob sends the instruction command to the car transceiver which then processes the requested instruction (lock/unlock car or open boot).

A successful transaction between car transceiver and key fob for starting the car is shown in Fig. 4. When the driver presses the start button it will send a request to the key fob to reply with the car key ID stored in the memory of the key fob. The key fob reads the car key ID, generates ten random numbers in the range between 0 and 1999 and sends the car key ID together with the ten generated random numbers to the car transceiver. After receiving the car key ID, the car transceiver checks if the received car key ID matches with the own one. If the comparison was successful, then the car transceiver builds, based on the received random numbers, the authentication message. Furthermore, it generates ten random numbers  in the range between 0 and 1999 before sending the authentication message together with the ten generated random numbers to the key fob. After receiving the authentication message, the key fob builds its own authentication message and compares it with the received one. If the comparison was successful, the key fob will build the authentication message based on the random numbers received from the car transceiver and sends it to the car transceiver. The car transceiver builds its own authentication message and compares it with the received one. In case both authentication messages match, the start command will be processed. It is to mention, that the authentication process (A) is necessary to avoid that an attacker emulates a car transceiver in order to get access to the memory locations of the key fob.

\begin{figure}[!ht]
\centering
\includegraphics[width=3.3in]{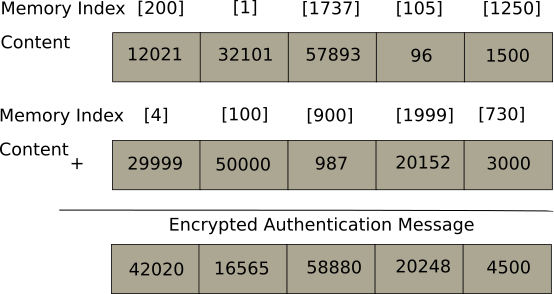}
\caption{The proposed Lightweight Encryption Algorithm.}
\label{fig_Lightweight_Encryption_Algorithm}
\end{figure}

The numbers that have been read from the requested memory locations can be encrypted using the proposed Lightweight Encryption Algorithm (see Fig. 5) that works in the following way. First the number of the sixth requested memory location is added to the number of the first requested memory location. Then the number of the seventh requested memory location is added to the number of the second requested memory location. This continues until the number of the tenth requested memory location is added to the number of the fifth requested memory location. This kind of encryption has the advantage that it needs less computation power and it is easy to implement. Fig. 6 illustrates the key exchange between the car transceiver and the two key fobs. As mentioned before the keys must be updated at certain time intervals.

To exchange the keys (random numbers), both key fobs must be connected to the hardware interfaces in the central console. After entering the password of the board computer, the key fob programming button becomes visible. By pressing the key fob programming button, the board computer will send a car key ID request to both car keys. If both car key IDs match, then the board computer generates 2000 random numbers and transmits them to both key fobs where they are stored in the EEPROM memory. If the random numbers have been successfully saved in both key fob memories then the board computer writes  the random numbers to its own EEPROM memory.

\begin{figure}[!ht]
\centering
\includegraphics[width=3.2in]{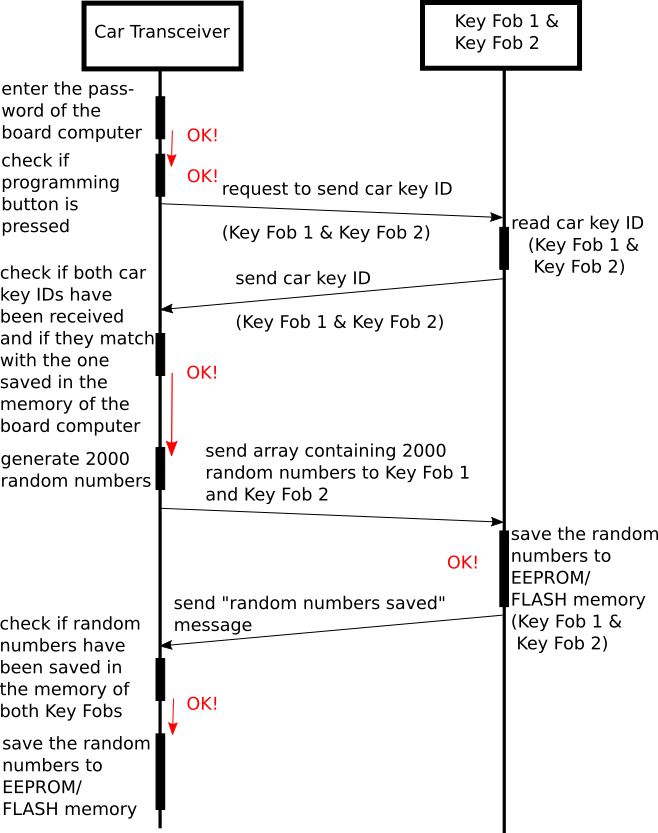}
\caption{Successful Transaction between car transceiver and Key Fob  for \noindent\hspace*{10mm} exchanging random numbers.}
\label{fig_Random_Numbers_Exchange}
\end{figure}

In the previous described transactions between car transcei-\\ver and key fob(s) it was always assumed that all the steps in the transaction went fine.
However some steps can fail. It can happen that the car key ID does not match, the comparison of the received encrypted message fails or a memory problem occurs when updating the key fobs with new keys (random numbers). These problems can be handled in the following way. If a wrong car key ID is received many times within a short time interval, it can be ignored because on a big parking yard there might be many cars which are opened and closed.
However, when receiving a wrong encrypted authentication message from a key fob many times within a short time interval, then the risk of an Intrusion Attack is very high because this step is only reached when the right car key ID was sent. In this case the system could be put in a blocked state after receiving three wrong authentication messages for three minutes, before it reacts to new key fob commands.

The memory problem that can occur when updating the key fobs with new keys (random numbers) can be solved in the following way. If a writing process of the keys to one of the key fobs fails then it should be repeated. In case it still fails, then the writing process should be stopped but only if the other key fob has not been programmed yet. Otherwise, the already programmed key fob must be re-programmed with the old keys.

\begin{figure}[!ht]
\centering
\includegraphics[width=3.3in]{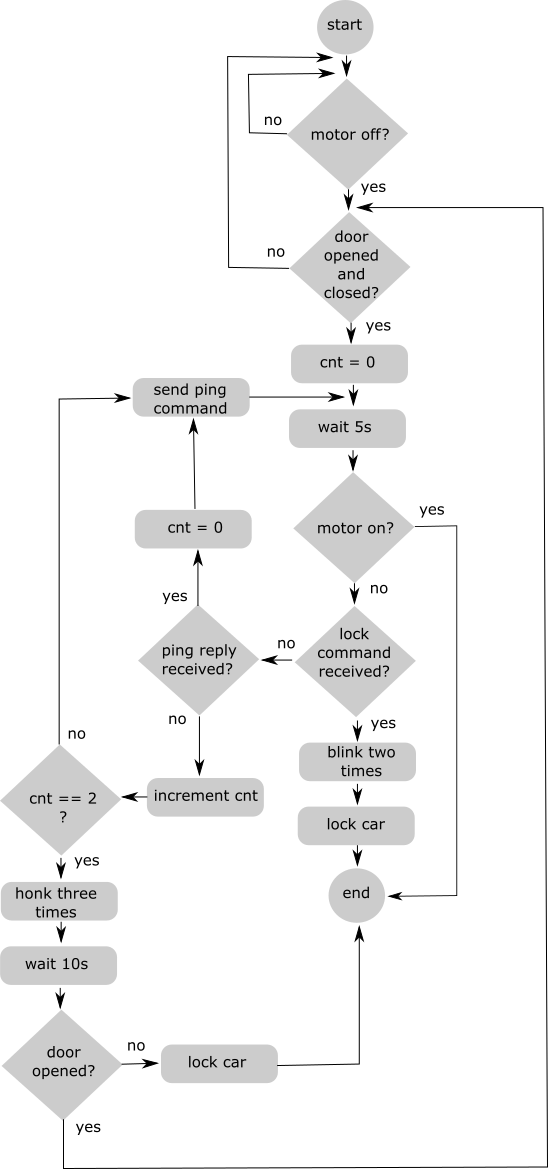}
\caption{A proposed security mechanism against jamming.}
\label{fig_security_mechanism_against_jamming}
\end{figure}

In section one, the operation of a jammer has been introduced. Fig. 7 represents the flow chart of the proposed security mechanism against jamming. After the motor is switched off the board computer will check if a door has been opened and closed. In case a door was opened and closed the board computer will check if a ping reply or a lock command has been received. If no ping reply or lock command has been received within ten seconds then the car will honk five times and the driver has ten seconds time before the car gets locked automatically. This prevents the car of being unlocked after a Jamming Attack.

\section{Comparison between different Authentication Techniques applied in Key Fobs}
For Remote Keyless Entry Systems there are three authentication techniques that are mainly used [3]. 

One technique is called Fixed Code Technique. A device, using this technique has a fixed pre-programmed code. Whenever an event is triggered, the device sends its fixed code to the receiver.

Another technique used for the authentication of Remote Keyless Entry Systems is the Rolling Code Technique. In this technique both, the transmitter (key fob) and the receiver (car) maintain a sequence counter. If a button is pressed, then the content of the sequence counter is sent encrypted to the receiver. There, the received encrypted value of the sender's sequence counter is decrypted and compared with the receiver's own sequence counter. If the difference between the values of both sequence counters is in a certain range, the received code is valid. It is to mention, that a shared secret key is used for the encryption and decryption.

The Challenge-Response Technique is a widely used technique. A secret key is shared between two transceivers. In case the driver pulls the door handle, the car transceiver transmits a random number, the so called random challenge, to the CID. There the received random challenge is going to be encrypted before it is sent back to the car transceiver. In the meanwhile the car transceiver has also encrypted the random challenge sent to the CID and compares it with the received one. If they match a certain operation is performed.

According to [3], the previous mentioned authentication techniques are vulnerable against different types of attacks such as Scan Attack, Playback Attack and Forward Prediction Attack. Table I shows the level of success of each attack type against the mentioned authentication techniques and against our propsed protocol. The levels are easy (0), difficult (1), very difficult (2) and extremely difficult (3).

\begin{table}[!ht]
\caption{Level of different Attacks vs. Authentication Techniques}
\label{table_attackLevel}
\centering
    \begin{tabular}{ | p{2.3cm} | p{0.8cm} | p{1cm} | p{1.2cm} | p{1.2cm} | }
    \hline \textbf{} & \textbf{Fixed \newline Code} & \textbf{Rolling \newline Code} & \textbf{Challenge \newline Response} & \textbf{Our \newline proposed      
    \newline Protocol} \\ 
    \hline \textbf{Scan Attack} & 1 & 1 & 2 & 3 \\
    \hline \textbf{Playback Attack} & 0 & 1 & 2 & 3 \\
    \hline \textbf{Fwd. Pred. Attack} & 0 & 1 & 2 & 3 \\
    \hline
    \end{tabular}
\end{table} 

\section{Benefits and Drawbacks of the proposed Protocol}
The introduced protocol has several advantages among common protocols used in Secure Remote Keyless Entry Systems. One advantage is the easy implementation of the introduced protocol and of the lightweight encryption algorithm. In addition the lightweight encryption algorithm requires less computation power and thus it is energy efficient. The authentication response message is built from three randomly selected decimal numbers each in the range between 0 and 65535. This means that the message length can be up to 80 bits long. Thus it is almost impossible to guess the requested authentication response message. Since the authentication response message changes for every request, the probability of guessing the right authentication response message is \begin{math} \frac{1}{2^{80}} \end{math}.
Another big advantage is that if someone borrows  the car to make a test drive with the purpose of cloning a car key, the person might not be able to steal the car in the night from the salesman's yard, if the salesman has re-programmed the key fobs after the test drive.

In the proposed Remote Keyless Entry System, the driver has to press a button on the key fob to open or lock the car. This prevents the
system from a Two-Thief Attack. Furthermore, with the proposed system it is almost impossible to become a victim of a Scan, Playback or a Challenge Forward Prediction Attack, since the authentication message consists of encrypted decimal numbers from randomly selected memory locations.

For a successful Scan Attack, the 32 bit car key ID, the always changing 80 bit authentication message and the timing must be correct before sending the instruction command to unlock the car. Hence, it is extremely difficult to attack the proposed protocol with a Scan Attack. 

For a successful Playback Attack, the attacker must hope that the recorded sequences of ten random numbers, used to access the values of ten memory locations in order to build the authentication message, will occur again. Since RNGs are used, the probability of repeated random number sequences is very small and thus it is extremely difficult to attack  the proposed protocol with a Playback Attack.

Attacking the proposed protocol with a Forward Prediction Attack is extremely difficult because the authentication message is built from the values of ten randomly selected memory locations. The indices of that memory locations are generated with a RNG and thus it is almost impossible to predict the sequence of the generated random numbers. 

One drawback of the proposed protocol is, that the owner is responsible for updating the key fobs with new keys (random numbers). When the board computer displays a request for updating the key fobs then the owner has to plug both key fobs into the interfaces located at the central console, so that they can be  programmed with the new keys. How often new keys should be written to the key fobs depends on how often the key fob(s) is/are used or at latest after a certain time interval.

\section{Conclusion}
In this paper we have introduced a Protocol for a Secure Remote Keyless Entry System applicable in Vehicles. For the encryption, symmetric cryptography is used. The system consists of a car transceiver and a key fob with programming interface. To process a command like unlock the car, the transceiver first requests the key fob to authenticate. This authentication is done by requesting the key fob to send an authentication message that is built from the content of ten randomly selected memory locations. The car transceiver builds its own authentication message and compares it with the received one. It is assumed that key fob as well as the car transceiver have an EEPROM where the shared keys are stored. To synchronize the content of the EEPROM of key fob and car transceiver, the programming interface is used. In this paper we have also compared our proposed protocol with existing authentication techniques applied in key fobs. We can conclude that the probability of hacking our proposed protocol with a Scan Attack, Playback Attack, Forward Prediction Attack or a Two-Thief Attack is extremely difficult. Furthermore, due to the lightweight encryption algorithm the processors in the car transceiver and in the key fobs need less computation power and thus the system becomes more energy efficient. More optimization techniques and mathematical proofs are left for the future work.


\end{document}